\documentstyle[12pt,epsf]{article}
\newcommand{\nll}{\nonumber \\}

\newcommand{\bq}{\begin{equation}}
\newcommand{\eq}{\end{equation}}
\newcommand{\ba}{\begin{eqnarray}}
\newcommand{\ea}{\end{eqnarray}}
\def \3{\ss}

\begin{document}
\voffset -2cm
\vspace{-2.5cm}
\begin{flushleft}
LMU-16/01
\end{flushleft}
\begin{center}
\vspace{1.5cm}\hfill\\ 
{\LARGE Why extra gauge Bosons should exist\\ and how to hunt them}\footnote{
talk presented at the symposium ``100 Years Werner Heisenberg -- Works and 
Impact'', Bamberg, September 2001, organized by the Alexander 
von Humboldt Foundation}
\vspace{1cm}\\
A. Leike,\smallskip\\
{\it
Ludwig--Maximilians-Universit\"at, Sektion Physik, Theresienstr. 37,\\
D-80333 M\"unchen, Germany}\\
E-mail: leike@theorie.physik.uni-muenchen.de\\
\end{center}

\begin{abstract}
Werner Heisenberg's work is the foundation for many topics of present research.
This is also true for the search for extra gauge bosons. 
The prospects of future colliders in this search are shortly mentioned.
\vspace{1cm}
\end{abstract}
%
%

During the last years of his life, Werner Heisenberg tried to found a unified
theory of all interactions.
Today, 25 years later, this unified theory is still lacking.
The minimum unification group of strong and electroweak interactions is 
$SU(5)$ \cite{su5}.
It is ruled out by several experiments. LEP has measured \cite{PDB}
$\sin^2\hat\theta(M_Z)(\overline{MS})=0.23117(16)$, while $SU(5)$ predicts
$\sin^2\theta(M_Z)=0.21$.
Proton decay is not observed in experiment \cite{PDB} leading to 
$\tau (p\rightarrow e^+\pi^0)>1.6\cdot 10^{33}$ years, while $SU(5)$ predicts
$\tau (p\rightarrow e^+\pi^0)=10^{28} - 10^{30}$ years.
The observed neutrino mixing in solar and atmospheric neutrino experiments 
indicates that this particle must have a non-zero mass, which is not the case 
in $SU(5)$.

The next extension of the unification group is $SO(10)$ \cite{fritzsch}.
It is consistent with experimental data.
All unification groups larger than $SU(5)$ contain extra neutral and extra 
charged gauge bosons.
Experimental signals of these particles would give interesting information 
about the underlying grand unified theory.
The search for extra gauge bosons is therefore a part of the physical programme
of every present and future collider.

In the vacuum of a quantum field theory, virtual particles are continuously 
emitted and absorbed. 
The energy needed to create these virtual particles can 
be ``borrowed'' from the vacuum due to the tunnel effect. 
However, these effects happen at length scales of the order of the de 
Broglie wavelengths of the involved particles.
Therefore, one must resolve
very short distances to be sensitive to heavy particles.
This can be done by high energy experiments or by precision experiments.

The higher the energy of an particle the higher is its momentum.
According to Heisenberg's uncertainty relation \cite{uncert}, test particles
with higher momentum can resolve smaller scales.
This is well known from electron microscopes.

High precision measurements can also give valuable information on heavy 
particles because they are sensitive to rare processes.
Consider, for example, atomic parity violation.
In the conservative atomic model, one assumes that only photons are exchanged 
between the atomic nucleus and the electrons. 
However, the interaction between an electron and a 
quark of the nucleus can also be mediated by the $Z$ Boson.
This is a rare process because the $Z$ Boson is heavy.
Photon exchange does not violate parity but $Z$ Boson exchange does.
Parity violating transitions are observed in experiments.
They prove quantitatively that $Z$ Bosons {\it are} exchanged in atoms.
The atomic parity violation experiments are so precise that they also set 
limits on extra neutral gauge bosons, although these particles must be 
considerably heavier than the $Z$ \cite{PDB}.

Extra gauge bosons are not observed. This means that they are either not 
existing or heavy. 
If extra gauge bosons are not too heavy they will show up in future 
experiments.
I will concentrate here on limits on extra gauge bosons from
a future linear $e^+e^-$ collider and compare them with possible limits from
LHC.
The expected experimental signal of these particles in future experiments
is a small deviation of observables from the Standard Model prediction. 
Future experiments will either see deviations from the SM or not.
In the first case it must be decided whether these deviations can be due to 
extra gauge bosons. If yes, the corresponding constraints on their parameters 
must be found.
In the second case, the data can be interpreted as exclusion limits to 
extra gauge bosons.
In both cases, it is necessary to calculate the measured observables
in theories with extra gauge bosons.
A recent review about these calculations for extra neutral gauge bosons
can be found in \cite{habil}.

Collision experiments are calculated with the help of the S-matrix.
Heisenberg founded the S-matrix theory in 1943 \cite{smatrix}.
The results of recent investigations on the search potential for extra gauge 
Bosons at a future linear $e^+e^-$ collider can be found in the TESLA 
technical design report \cite{tdr}.

The discovery limits for extra gauge bosons are shown in tables~1 and 2. 
The numbers in the tables also show the dependence of the limits on the 
expected systematic errors.
Case A and B in table 1 refer to the following assumptions on systematic
errors \cite{srlcnote}:
\ba
\mbox{case\ A:\ \ }\Delta P_{e^\pm}=1.0\%,\ \Delta L = 0.5\%,\ 
\Delta^{sys}\epsilon_{lepton}=0.5\%, \Delta^{sys}\epsilon_{hadron}=0.5\%\nll 
\mbox{case\ B:\ \ }\Delta P_{e^\pm}=0.5\%,\ \Delta L = 0.2\%,\ 
\Delta^{sys}\epsilon_{lepton}=0.1\%, \Delta^{sys}\epsilon_{hadron}=0.1\%
\ea
The beam polarizations are $P_{e^+}=0.6$ and $P_{e^-}=0.8$ in both tables.

The numbers in table 2 are sensitive not only to systematic errors but also
to kinematical cuts, which take into account the detector acceptance.
These cuts are also needed to suppress the SM background and to remove singularities arising due to soft or collinear photons.
The limits from $e\gamma$ collisions assume backscattered laser photons.
See references \cite{alwp} for details.

As we see from the tables, hadron and lepton colliders are 
complementary in a  search for extra gauge bosons.
In the case of a positive signal, a lepton collider is especially strong in 
model discrimination.
See references \cite{srlcnote, alwp} for a discussion of that point. 

\begin{table}[t]
\begin{center}
\begin{tabular}{|l|lllll|l|}
\hline
&\multicolumn{2}{c}{$\sqrt{s}=0.5\,TeV$,}&
 \multicolumn{2}{c}{$\sqrt{s}=1\,TeV$,} &&
LHC,\\
&\multicolumn{2}{c}{$L_{int}=1000fb^{-1}$} &
 \multicolumn{2}{c}{$L_{int}=1000fb^{-1}$} && 
$L_{int}=100fb^{-1}$\\
&\multicolumn{2}{c}{$e^+e^-\rightarrow f\bar f$}
&\multicolumn{2}{c}{$e^+e^-\rightarrow f\bar f$}&&\\
case (syst.err.)& A & B & A & B &&\\
\hline
Model &&&&&&\\ \hline
$\chi$& 6.2 & 6.9 &  9.5 & 12.2 && 4.3 \\
$\psi$& 2.1 & 3.4 &  4.0 &  5.5 && 4.2 \\
$\eta$& 3.3 & 4.0 &  5.3 &  6.6 && 4.3 \\
$LR$  & 7.2 & 9.3 & 13.7 & 15.6 && 4.6 \\
\hline
\end{tabular}
\end{center}
{\bf Table 1:} {\it Discovery limits on extra neutral gauge bosons in $TeV$ 
for an $e^+e^-$ collider.
The numbers are taken from Fig. 5.2.1. of \cite{tdr}. 
See the original reference \cite{srlcnote} for more details.}
\end{table} 

\begin{table}[t]
\begin{center}
\begin{tabular}{|l|lllllllll|}
\hline
&\multicolumn{4}{c}{$\sqrt{s}=0.5\,TeV, L_{int}=1000fb^{-1}$} &
 \multicolumn{4}{c}{$\sqrt{s}=1\,TeV,   L_{int}=1000fb^{-1}$} &\\
&\multicolumn{2}{c}{$e^+e^-\rightarrow \nu\bar\nu\gamma$}
&\multicolumn{2}{c}{$e\gamma\rightarrow \nu q + X$} &
 \multicolumn{2}{c}{$e^+e^-\rightarrow \nu\bar\nu\gamma$}
&\multicolumn{2}{c}{$e\gamma\rightarrow \nu q + X$} &\\
syst.err,\%&0.1&2.0&0.5&2.0&0.1&2.0&0.5&2.0&\\
\hline
Model &&&&&&&&&\\ \hline
SSM $W'$& 4.8 & 1.7 & 4.0 & 2.7 & 5.9 & 2.2 & 5.8 & 4.6&\\
LRM     & 1.3 & 0.9 & 0.7 & 0.6 & 1.7 & 1.2 & 1.2 & 1.1&\\
KK      & 5.0 & 1.8 & 5.7 & 3.8 & 6.4 & 2.3 & 8.2 & 6.5&\\
\hline
\end{tabular}
\end{center}
{\bf Table 2:} {\it Discovery limits on extra charged gauge bosons in $TeV$
for an $e^+e^-$ collider.
The numbers are selected from table 5.2.1. of \cite{tdr}.
See the original references \cite{alwp} for more details.
The discovery limits for LHC, $L_{int}=100fb^{-1}$ are between 
5.1 and 5.9\,TeV \cite{rizzo}
}
\end{table} 


\bigskip
 
\noindent{\Large\bf Acknowledgement}\smallskip

I would like to thank the Alexander von Humboldt Foundation for inviting me 
to this perfectly organized stimulating conference.

\vfill
\end{document}